\documentclass[twocolumn,pre,superscriptaddress,floatfix]{revtex4}

\usepackage{graphicx}
\usepackage{subfigure}
\usepackage{amsmath}
\usepackage{amssymb}
\usepackage{bm}
\usepackage{times}

\usepackage[pdftex]{hyperref}
\hypersetup{colorlinks=true,linkcolor=blue,citecolor=blue,urlcolor=blue}

\newcommand{\pbc}{\pi}
\newcommand{\abc}{\overline{\pi}}

\begin{document}

\title{Ground state interface exponents of the diluted Sherrington-Kirkpatrick spin glass}

\author{Wenlong Wang}
\email{wenlongcmp@scu.edu.cn}
\affiliation{College of Physics, Sichuan University, Chengdu 610065, China}

\begin{abstract}
We present a large-scale simulation of the ground state interface properties of the diluted Sherrington-Kirkpatrick spin glass of Gaussian disorder for a broad range of the bond occupation probability $p$ using the strong disorder renormalization group and the population annealing Monte Carlo methods. We find that the interface is space-filling independent of $p$, i.e., the fractal dimension $d_s=1$. The stiffness exponent $\theta$ is likely also independent of $p$, despite that the energy finite-size correction exponent $\omega$ varies with $p$ as recently found. The energy finite-size scaling is also analyzed and compared with that of the $\pm J$ disorder, finding that the thermodynamic energy is universal in both $p$ and the disorder, and the exponent $\omega$ varies with $p$ but is universal in the disorder.
%and the prefactor has no universality.
%Our work suggests that the diluted and the full Sherrington-Kirkpatrick spin glasses are likely in the same zero-temperature universality class.
\end{abstract}

%\pacs{75.50.Lk, 75.40.Cx, 05.50.+q}

\maketitle

\section{Introduction}

Spin glasses are disordered and frustrated magnets with numerous intriguing properties, e.g., the unusual replica symmetry breaking (RSB) \cite{parisi:79,parisi:80,parisi:83}, equilibrium temperature and bond chaos \cite{Hukushima:Chaos,Katzgraber:Chaos,Wang:EA4D,Zhai:TC,Berker:Chaos}, and various exotic nonequilibrium dynamics such as hysteresis, memory and rejuvenation effects \cite{fisher:91d,sales:02,silveira:04,Doussal:Chaos}. While the short-range Edwards-Anderson (EA) spin glass \cite{EA} in finite dimensions remains controversial, spin glass physics has already found applications in diverse fields such as optimization problems \cite{Haijun:EI,book}, neural networks \cite{Young:RMP,book}, and structural glasses \cite{charbonneau:14,Mike:GT}.  As such, spin glass is frequently referred to as a prototypical complex system, a theme of the 2021 Nobel prize. 

A central question in spin glass physics is what properties of the mean-field Sherrington-Kirkpatrick (SK) spin glass \cite{SK} are inherited by the short-range EA spin glass. The RSB picture \cite{parisi:08,mezard:87} assumes that the mean-field description is qualitatively correct for the EA model, there are many pure states and droplet interfaces are space-filling. On the other hand, the droplet picture \cite{mcmillan:84b,bray:86,fisher:86,fisher:87,fisher:88} based on the domain-wall renormalization group method predicts only a single pair of pure states like a ferromagnet and the interfaces are fractals below an upper critical dimension, see, e.g., \cite{Wang:Fractal,Wang:Fractal2}. The two pictures also disagree on the existence of a spin-glass phase in a weak external magnetic field \cite{almeida:78}, and there are other pictures as well \cite{book}. Numerical results suggest many pure states but a fractal dimension in three dimensions, but the interpretations remain unsettled due to the limited range of sizes, see, e.g., \cite{Mike:replicon} for a recent discussion.

To gain more insight on the EA spin glass, it is helpful to study spin glass models that deviate from the SK model but not as dramatically as that of the EA model. 
The most intensively studied model towards this direction is arguably the so-called Bethe lattice \cite{mezard:01}, where each spin has a finite connectivity or degree $z$, either fixed or on average, on a random graph. The one-dimensional Kotliar-Anderson-Stein (KAS) spin glass \cite{kotliar:83} with power-law interactions, and also the $p$-spin models \cite{Guerra2003} have also attracted considerable attention. Recently, the bond diluted SK model was studied \cite{Boettcher:SKD}. This model differs from the Bethe lattice in that $z=pN$ for $N$ spins, where $p$ is the bond occupation probability. Interestingly, this model had received little attention, presumably because at first sight this model should behave similarly as the SK model since $z$ eventually diverges. However, it was recently found that the energy finite-size correction exponent $\omega$ varies nontrivially with $p$ \cite{Boettcher:SKD}. This naturally raises an important question when and how is this parameter $p$ relevant. Particularly, it is highly interesting whether the interface exponents, i.e., the stiffness exponent $\theta$ and the fractal dimension $d_s$ depend on $p$. These two exponents are of paramount importance for understanding the interface properties \cite{Wang:KAS,Wang:Fractal}. In addition, they are also used in characterizing the spin-glass chaos \cite{Hukushima:Chaos,Katzgraber:Chaos}. 

Our motivation is that the energy finite-size correction exponent $\omega$ (see Eq.~\eqref{ps}) appears to be correlated with the stiffness exponent $\theta$ for some models, e.g., $\omega=1-\theta/d$ was proposed for the EA model \cite{Boettcher:GS}. For the diluted SK model, the exponent $\omega$ appears to increase rather violently as $p \rightarrow 0$, suggesting that $\theta$ may decrease with decreasing $p$. If $\omega$ diverges to infinity in the $p=0$ limit, then the formula $\omega=1-\theta$ seems to suggest that $\theta<0$ for a sufficiently small $p$ if a similar argument is applied to the diluted SK model. It was found that $\omega \gtrsim 1$ when $p \lesssim 0.03$, e.g., $\omega=1.32(1)$ at $p=0.01$ \cite{Boettcher:SKD}. 
However, $\theta<0$ is unreasonable, as it would imply an absence of the spin-glass phase when $p$ is small. Therefore, the strong dependence of $\theta$ on $p$ may not exist, contrary to the exponent $\omega$. Nevertheless, it is still interesting whether a weak correlation exists. On the other hand, it seems likely that the interfaces are space-filling, i.e., $d_s=1$ for the diluted SK model independent of $p$ because this is a mean-field model on a random graph.

The main purpose of this work is to systematically study the interface exponents $\theta$ and $d_s$ as a function of $p$ for the diluted SK spin glass. Here, the interface of the SK model is studied as the long-range limit of the one-dimensional KAS model. The exponents are studied using two different methods for a more coherent understanding. The $d_s$ exponent is studied to very large sizes to reduce systematic errors using the strong disorder renormalization group (SDRG) method, which is a heuristic but remarkably effective method for estimating $d_s$ \cite{Wang:Fractal,Wang:Fractal2}. This method, however, cannot reliably estimate $\theta$. To this end, we find numerically exact ground states using population annealing Monte Carlo (PAMC) simulations. Both the $d_s$ and $\theta$ exponents are analyzed in this framework. Our result suggests that the interfaces are all space-filling, and $\theta$ has no or alternatively very weak dependence on $p$.

%In fact, our simulations are equilibrium simulations, and finite-temperature data are also available, despite that our primary focus here is the ground state. \cite{Wang:Fractal2}.

%Here, a pure state refers to a self-sustained thermodynamic state characterized by a time-averaged spin orientational pattern. The RSB picture \cite{parisi:08,mezard:87} assumes that the mean-field theory is qualitatively correct for the EA model. On the other hand, the droplet picture \cite{mcmillan:84b,bray:86,fisher:86,fisher:87,fisher:88} based on the domain-wall renormalization group method predicts only a single pair of pure states much like a ferromagnet. The two pictures also differ on the geometrical aspect of excitations (fractals or space-filling) \cite{Wang:Fractal}, and the existence of a spin-glass phase in a weak external magnetic field \cite{almeida:78}. There are other pictures as well \cite{book}. The applicability of RSB is of broad interest and is related to, e.g., the Gardner transition in structural glasses \cite{charbonneau:14,Mike:GT}.

%\cm{The central questions now are how $\theta$ and $d_s$ vary with $p$. Particularly, can we get a $p$-dependent $\theta$ and then a $p$-independent $d_s$? If so, our results would be highly interesting. The models provide a family of models with tunable stiffness and thereby chaos.}

An additional purpose of this work is to compare the scaling of the ground state energy per spin with \cite{Boettcher:SKD} as we used Gaussian disorder instead of the $\pm J$ disorder. First, it is theoretically expected that the thermodynamic energy per spin is independent of the specific types of disorder \cite{Univ:SK}, likely also independent of $p$. Indeed, our work suggests that the thermodynamic average energy per spin is universal in both $p$ and the disorder. The exponent $\omega$ is universal in the disorder, but not in $p$ as mentioned earlier. Finally, the scaling prefactor has no universality.

%Second, it is interesting whether the exponent $\omega$ is the same for the two models. These quantities are also briefly discussed in this work.
%it is already quite evident that $A$ is not universal. We need to study $p=0.1$ well and see if we get the same $\omega$ within errorbars.

This work is organized as follows. In Sec.~\ref{setup}, we introduce the model and the numerical setup. Next, we present our numerical results in Sec.~\ref{results}. Finally, our conclusions and some future considerations are summarized in Sec.~\ref{conclusion}.

%It might be tempting to consider that this model is essentially the same as the SK model for any finite $p$, explaining why it was not extensively studied. However, it was shown recently that the energy finite-size correction exponent $\omega$ interestingly depends nontrivially on $p$ by studying the finite-size scaling of the ground state energy per spin. This presumably suggests that the exponent $\theta$ depends nontrivially on the parameter $p$, $\omega=1-\theta$? But $\omega$ runs above $1$ if $p \lesssim 0.03$, absence of a spin-glass phase or breakdown of $\omega = 1-\theta$?

%If they are all space-filling, then maybe we can continue to study $d_s$ as a function of $\sigma$ for the KAS model. Is this the same as the one of $p=1$?

\section{Computational setup}
\label{setup}

The SK model has no geometric structure, therefore, the SK interface is frequently viewed as the infinite-range limit of the one-dimensional KAS model with power-law interactions \cite{kotliar:83}. In the KAS model, the interaction strength between two spins separated by a distance $r$ decreases proportional to $1/r^\sigma$ where $\sigma>0$ is an exponent controlling the interaction range. The KAS model has attracted considerable attention in its own right because it effectively interpolates between the $d=1$ and $d=\infty$ of the EA
model \cite{bray:86b,fisher:88,katzgraber:03}. Renormalization group arguments deduce that the model is expected to behave like the infinite-range SK model for $0 \le \sigma <1/2$, and for $1/2< \sigma
<2/3$ the critical exponents at the spin-glass transition are mean-field
like and this corresponds to the EA model with space dimensions above
six. Below $\sigma=2/3$, the model deviates from the mean-field regime. Both $\theta$ and $d_s$ are expected to be independent of $\sigma$ in the entire mean-field regime $\sigma<2/3$, and numerical works have confirmed this independence in the regime $\sigma \lesssim 0.2$ \cite{bray:86b,Wang:KAS,Wang:Fractal} due to finite-size effects. Here, we focus on this regime $\sigma \lesssim 0.2$. Next, we introduce bond dilution to the KAS model.

The one-dimensional KAS model with a bond occupation probability $p$ is described by the Hamiltonian: 
\begin{equation}
H = - \frac{1}{\sqrt{p}} \sum_{i<j} \epsilon_{ij} J_{ij} S_i S_j,
\label{Hamiltonian}
\end{equation}
where $L$ Ising spins $S_i=\pm 1$ lie on a ring, and the exchange interaction $J_{ij}$ is occupied when $\epsilon_{ij}=1$ with probability $p$ or otherwise the bond is unoccupied when $\epsilon_{ij}=0$, $J_{ij}$ is defined as:
\begin{equation}
J_{ij}=c(\sigma,L) \frac{\zeta_{ij}}{r_{ij}^{\sigma}},
\label{eqn:Jij}
\end{equation}
where $r_{ij}=\min\left(|j-i|, L-|j-i|\right)$ is the shorter distance between sites $i$ and $j$. The disorder $\zeta_{ij}$ is chosen from the standard Gaussian
distribution $n(0,1)$. The exponent $\sigma \geq 0$ controls the range of interactions, and the
constant $c(\sigma,L)$ is fixed such that the
mean-field transition temperature $T_c^{\mathrm{MF}} = (\sum_j [(\epsilon_{ij}/\sqrt{p})^2 J_{ij}^2]_{\mathrm{av}})^{1/2} =1$, where
$[\cdots]_{\rm av}$ represents a disorder average. As $\epsilon_{ij}$ and $J_{ij}$ are independent, it is straightforward to show that $1/c^2=
\sum_{j \neq i} 1/r_{ij}^{2 \sigma}$. The sum is independent of $i$ due to the symmetry of the lattice, and interestingly the normalization constant $c$ here is identical to that of the full KAS model without dilution, as the dilution effect has already been compensated in the Hamiltonian by the factor $1/\sqrt{p}$. When $p=1$, the model restores the full KAS model. When $\sigma=0$, the model becomes the diluted SK model studied in \cite{Boettcher:SKD}. To reduce finite-size effects, we demand that the spins are almost surely connected by requiring $pL\geq20$, therefore, as $p$ decreases, the minimum size should appropriately increase to avoid disconnected samples. Having defined the diluted KAS model, we introduce the domain-wall interface in the next.

To generate an interface, we run each disorder sample twice one with the periodic boundary condition (PBC, ${\pbc}$) and the other with the anti-periodic boundary condition (APBC, ${\abc}$). The APBC is produced by flipping the sign of the bonds when the shorter paths go through the boundary in the middle of $S_1$ and $S_L$ \cite{Wang:KAS,Wang:Fractal}. We primarily focus on the interface of the ground states $S_i^{(\pbc)}$ and $S_i^{(\abc)}$. We employ two methods for finding ground states, the SDRG method for finding approximate solutions and the PAMC method for finding numerically exact solutions.

The SDRG is a remarkably effective method for estimating the exponent $d_s$, despite that it only finds approximate ground states \cite{weigel:dw,Wang:Fractal,Wang:Fractal2}. The SDRG result is not exact, but it is highly accurate. The main advantage of this method is that one can simulate very large system sizes to significantly suppress the finite-size effect, essentially removing the systematic error of small sizes. This is particularly important here as the SK or KAS models frequently have considerable finite-size effects \cite{Wang:KAS}. The SDRG runs a spin elimination process \cite{Cecile:Fractal,Wang:Fractal}. First, a criterion is used to find two spins which interact most strongly and also which spin of this pair should be eliminated. In this step, the relative orientations of these two spins are determined, and a spin and the bond between them is deleted. Next, the remaining bonds of the eliminated spin are transferred to the survival spin, and these bonds are suitably scaled by the sign of the removed bond. This process is then repeated until only two spins are left, where the ground state can be readily found. As the relative orientations of the spins are recorded, the full ground state up to the $Z_2$ symmetry can be constructed. A technical description of this algorithm is pretty elaborated and considering also that the discussion below does not require a detailed understanding of the algorithm, we refer the interested readers to the references for details \cite{Cecile:Fractal,Wang:Fractal}. The SDRG method, however, cannot reliably estimate $\theta$. To this end, we find numerically exact ground states using PAMC simulations. This also allows us to benchmark our SDRG result, and also provides more detailed data such as the overlap distribution at finite temperatures.

Population annealing is an efficient sequential Monte Carlo method for equilibrium sampling glassy systems with rugged energy landscapes \cite{Hukushima:PA,Machta:PA,Wang:PA,Weigel:PA,Amey:PA,Amin:PA,UPA}. This method has been extensively applied recently in large-scale spin glass simulations because it has several attractive features, e.g., it has several intrinsic equilibration measures, it is efficient, and massively parallel. Finding spin glass ground states with a good confidence using PAMC was studied in \cite{Wang:GS}. The criteria simultaneously require that a population of replicas maintains thermal equilibrium throughout the annealing process and also that the number of replicas with the minimum energy is sufficiently large, e.g., at least $10$ at the lowest temperature \cite{Wang:GS}. Then, the miminum energy state naturally is almost surely the true ground state.
  
Population annealing slowly cools a population of $R$ replicas starting from, e.g., random states at $\beta=0$ with alternating resampling and Metropolis sweeps until reaching the lowest temperature following an annealing schedule. In a resampling step from $\beta$ to $\beta'$, a replica $i$ is copied $n_i$ times according to its energy $E_i$ with the expectation number $\tau_i=\exp[-(\beta'-\beta) E_i]/Q$. Here, $Q=(1/R)\sum_i \exp[-(\beta'-\beta) E_i]$ is a normalization factor to maintain the population size approximately $R$. In our simulation, $n_i$ is chosen as either the floor or the ceiling of $\tau_i$ with suitable probabilities. After a resampling step, $N_S$ Monte Carlo sweeps are applied to each replica at the new temperature $\beta'$ to reequilibrate the population. Our equilibrium criterion is based on the replica family entropy $S_f = -\sum_i f_i \log (f_i)$, where $f_i$ is the fraction of replicas descended from replica $i$ of the initial population. We require $S_f \geq \log(100)$ at the lowest temperature for each individual sample \cite{Wang:PA,Wang:GS}. As mentioned above, we also require that the number of replicas with the minimum energy is at least $10$ to find the ground state.
If these criteria are not satisfied for a sample, we increase the amount of work, e.g., the population size or number of sweeps until they are satisfied. To benchmark our solver, we also compared our ground states with exact solutions by the exact enumeration, and the Branch and Bound algorithm using the spin glass server \cite{RRW10,sgserver} for small sizes $L \lesssim 30$ and $L \lesssim 50$, respectively.
%The preliminary simulation parameters are summarized in Table~\ref{table}, and unequilibrated instances were rerun with larger parameters until equilibration is reached.

Next, we introduce several observables defined from the ground states $S_i^{(\pbc)}$ and $S_i^{(\abc)}$, and their scaling properties with respect to the system size. These observables closely follow their full KAS model counterparts, but here they are systematically generalized to the diluted system. The link overlap \cite{hartmann:02,katzgraber:03f} is defined as:
\begin{equation}
q_{\ell} = \frac{\sum_{i<j} \epsilon_{ij} w_{ij} 
S_i^{(\pbc)}S_j^{(\pbc)} 
S_i^{(\abc)}S_j^{(\abc)} 
(2 \delta_{J_{ij}^{\pbc},J_{ij}^{\abc}} -1)}{\sum_{i<j} \epsilon_{ij} w_{ij}},
\label{linkoverlap}
\end{equation}
where $w_{ij}=1/r_{ij}^{2\sigma}$ is proportional to the variance of the bond \cite{katzgraber:03f,Wang:KAS}. We mention in passing that we also implemented a simper version replacing the denominator with the disorder-averaged expectation value $p\sum_{i<j}w_{ij}=pL/(2c^2)$, and found this version can have significant fluctuations and therefore should be avoided. By contrast, the weights of links in Eq.~\eqref{linkoverlap} are properly normalized for each individual sample.

The exponents $\theta$ and $d_s$ are then extracted from the following scaling relations:
\begin{eqnarray}
|\Delta E_{\rm{GS}}| &=& |E_{\rm{GS}}^{(\pi)}-E_{\rm{GS}}^{(\overline{\pi})}| \sim L^{\theta}, \label{scalingtheta} \\
\Gamma \equiv 1-q_{\ell} &\sim& \frac{2\Sigma^{\rm DW}}{pL(L-1)/2} \sim L^{d_s-1}, \\
N_I &\sim& L^{d_s},
\end{eqnarray}
where the number of islands
\begin{equation}
N_I= \frac{1}{4} \sum\limits_{i=1}^{L} (\tau_{i+1}-\tau_i)^2,
\end{equation}
where $\tau_i=S_i^{(\pi)}S_i^{(\overline{\pi})}$ and $\tau_{L+1}=\tau_1$. Here, an island is a sequence in which the $\tau_i$ are of the same sign. In the RSB region where $d_s =d =1$, the typical island size $L_0=L/N_I$ is of $O(1)$ independent of the system size, a
result which we obtained previously in the SK limit without dilution \cite{Wang:KAS}. For a fractal domain wall, $d_s<1$, and in general $0\leq d_s \leq 1$ \cite{Wang:KAS}. The scaling of $\Gamma$ is quite abstract and it is motivated by the short-range EA model, the $\Sigma^{\rm{DW}}$ is the interface size or the number of affected bonds, each spin has $L^{d_s}$ flipped bonds, therefore $\Sigma^{\rm{DW}} \sim L^{d_s+1}$ and $\Gamma \sim L^{d_s-1}$.

Finally, the energy finite-size scaling is also studied:
\begin{align}
e_L &= E_{\rm{GS}}/L, \\
e_L &= e_{\infty} + \frac{A}{L^{\omega}}.
\label{ps}
\end{align}
Here, either periodic or anti-periodic boundary conditions can be used. We compare the Gaussian disorder results with the $\pm J$ disorder ones \cite{Boettcher:SKD}, and therefore study the universality properties of  $e_\infty$, $A$, and $\omega$ with respect to $p$ and the disorder.

%and the two should be equivalent, and both $A$ and $\omega$ depend on $p$. We aim to compare

\section{Results}
\label{results}

\subsection{Space-filling interface $d_s=1$}

\begin{figure*}
\subfigure[]{\includegraphics[width=0.49\textwidth]{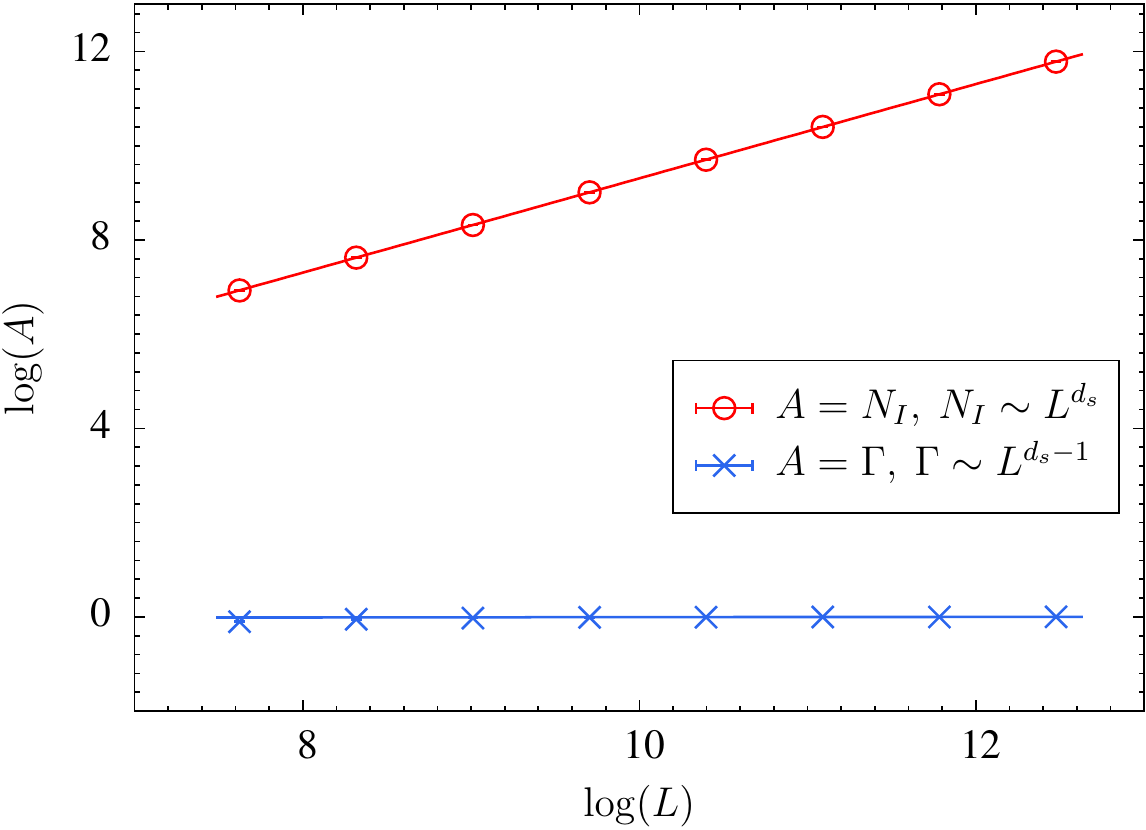} \label{P001}}
\subfigure[]{\includegraphics[width=0.49\textwidth]{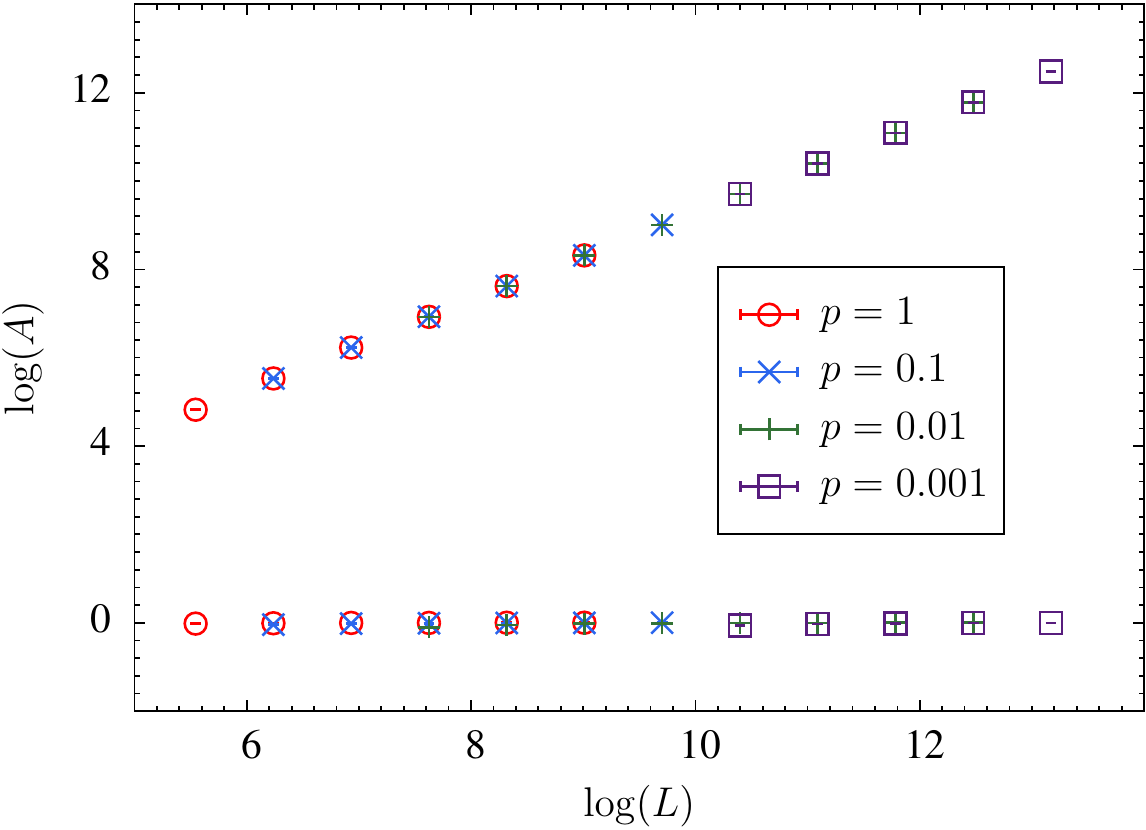} \label{SDRGSK}}
\subfigure[]{\includegraphics[width=0.49\textwidth]{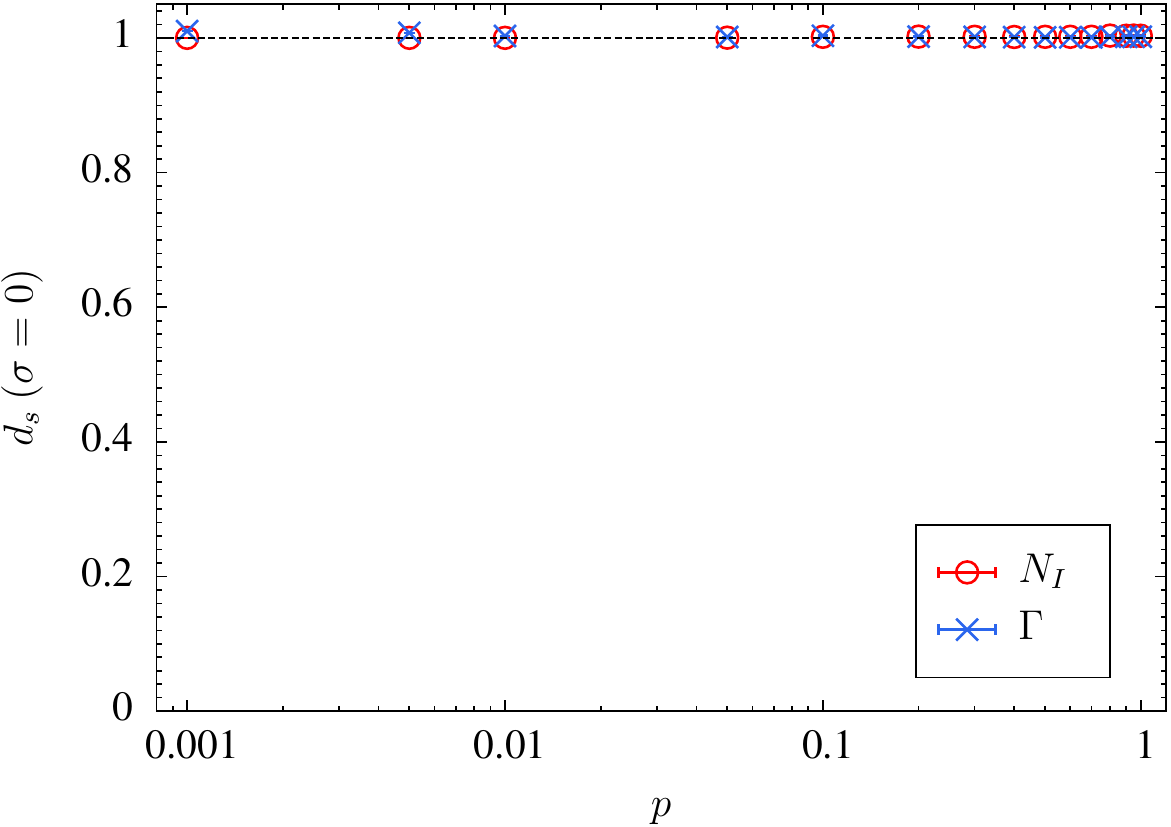} \label{ds}}
\subfigure[]{\includegraphics[width=0.49\textwidth]{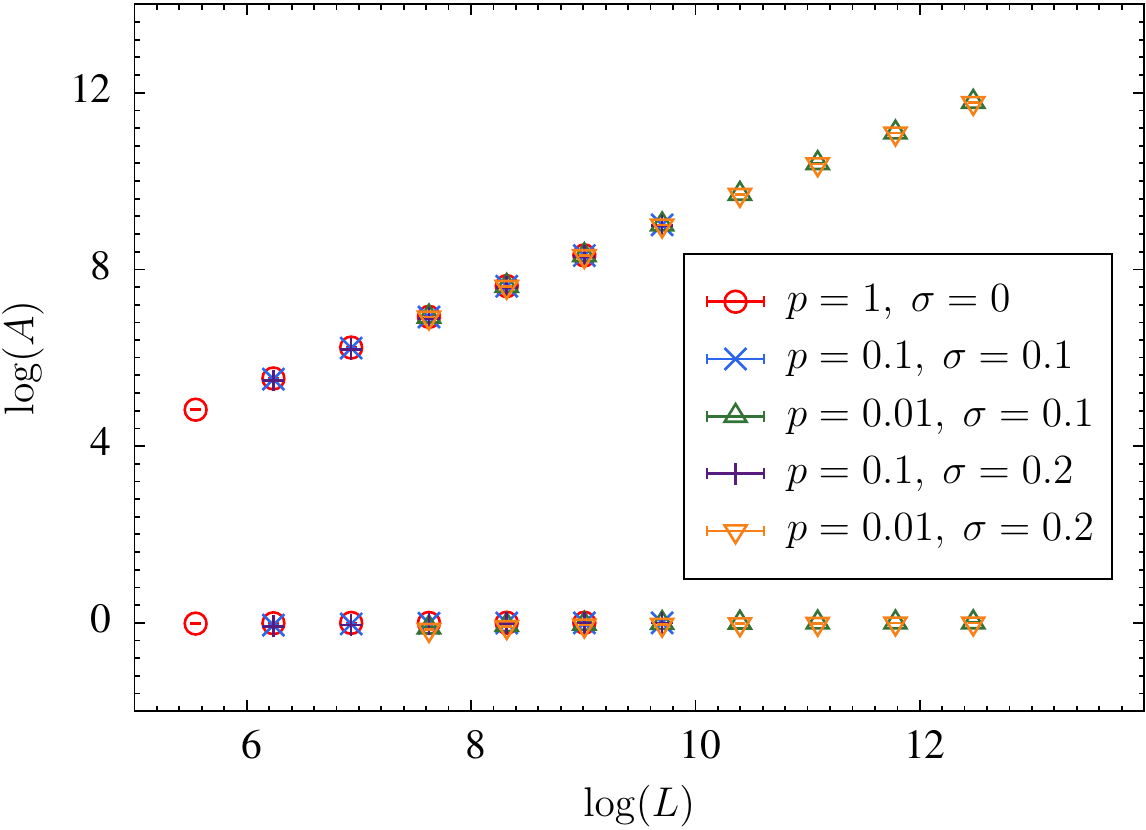} \label{SDRGKAS}}
\caption{
The SDRG results for the diluted SK model (a-c) and also the diluted KAS model (d). The first panel illustrates the scaling of both $N_I$ and $\Gamma$ for a typical case $p=0.01$ along with the linear fits using the five largest sizes (a). Both statistics yield approximately $d_s=1$. The second panel shows data at a few more values of $p$, and other data (not shown) also fall very well on the two lines, suggesting that the scaling functions are universal independent of $p$ (b). The third panel summarizes the extracted $d_s$ exponents from both statistics using the five largest sizes (c). The deviation of $d_s$ for $\Gamma$ at small values of $p$ is likely due to finite-size effect, see the slight bending of $\Gamma$ at small sizes in (a). The SDRG results suggest that the interface is space-filling independent of $p$. The last panel (d) suggests that both the scaling functions of $N_I$ and $\Gamma$ are universal in $p$ not only at $\sigma=0$, but extends further into the mean-field regime, here up to $\sigma=0.2$.
}
\label{SDRG}
\end{figure*}

The SDRG results coherently suggest that the interfaces are space-filling independent of $p$ and this also holds when $\sigma$ slightly departs from the SK limit $\sigma\leq 0.2$, as shown in Fig.~\ref{SDRG}.
We first discuss the diluted SK model in detail. Figure~\ref{P001} depicts the scaling of both $N_I$ and $\Gamma$ for a typical case $p=0.01$ along with the linear fits using the five largest sizes. Both statistics yield approximately $d_s=1$, the estimates are $1.00015(4)$ and $1.002973(3)$ from $N_I$ and $\Gamma$, respectively. The errorbars here are only statistical errorbars, and the tiny residual errors are largely systematic errors partly from the finite-size effect. Note that the $\Gamma$ curve slightly bends at small sizes. This downward curving at small sizes means that results from small systems tend to overestimate $d_s$. Indeed, if we had included the smaller sizes, the estimate would be yet slightly larger. This illustrates the major advantage of the SDRG approach, as one can reach very large sizes to suppress the finite-size effect. It is worth mentioning that this type of finite-size effect also exists for exact solutions, and the smallest size here $L=2048$ is a pretty daunting size for any exact method. The finite-size effect appears to be smaller for $N_I$ than for $\Gamma$, in line with \cite{Wang:Fractal}. The results strongly suggest that $d_s=1$ for $p=0.01$. The simulation parameters are summarized in Table~\ref{para}.

\begin{table}[t]
\caption{
Simulation parameters of the SDRG runs for the diluted SK and KAS models. Here, $\sigma$ is the power-law exponent of the KAS model, $p$ is the bond occupation probability, $L=2^K$ is the system size, and $M$ is the number of disorder realizations.
\label{para}
}
\begin{tabular*}{\columnwidth}{@{\extracolsep{\fill}} l l c r}
\hline
\hline
$\sigma$ &$p$ &$K$ &$M$ \\
\hline
$0$ &$\{1, 0.95, 0.9, 0.8\}$ &$\{8, 9, 10, 11, 12, 13\}$ &$5000$ \\
$0$ &$\{0.7, 0.6, 0.5, ..., 0.1\}$ &$\{9, 10, 11, 12, 13, 14\}$ &$5000$ \\
$0$ &$0.05$ &$\{10, 11, 12, 13, 14, 15, 16\}$ &$5000$ \\
$0$ &$0.01$ &$\{11, 12, 13, 14, 15, 16, 17\}$ &$5000$ \\
$0$ &$0.01$ &$\{18\}$ &$145$ \\
$0$ &$0.005$ &$\{12, 13, 14, 15, 16, 17\}$ &$5000$ \\
$0$ &$0.005$ &$\{18\}$ &$1000$ \\
$0$ &$0.001$ &$\{15, 16, 17, 18\}$ &$5000$ \\
$0$ &$0.001$ &$\{19\}$ &$1670$ \\
$0.1$ &$0.1$ &$\{9, 10, 11, 12, 13, 14\}$ &$5000$ \\
$0.1$ &$0.01$ &$\{11, 12, 13, 14, 15, 16\}$ &$5000$ \\
$0.1$ &$0.01$ &$\{17\}$ &$1728$ \\
$0.1$ &$0.01$ &$\{18\}$ &$250$ \\
$0.2$ &$0.1$ &$\{9, 10, 11, 12, 13, 14\}$ &$5000$ \\
$0.2$ &$0.01$ &$\{11, 12, 13, 14, 15, 16\}$ &$5000$ \\
$0.2$ &$0.01$ &$\{17\}$ &$1475$ \\
$0.2$ &$0.01$ &$\{18\}$ &$87$ \\
\hline
\hline
\end{tabular*}
\end{table}

The scaling functions appear to be universal in $p$, as illustrated in Fig.~\ref{SDRGSK}. Here, the data are shown for a few typical values of $p$, and they all fall approximately on the same lines independent of $p$. Particularly, the parameter $p$ here spans three orders of magnitude. The extracted exponents from both $N_I$ and $\Gamma$ are summarized in Fig.~\ref{ds} using again the five largest sizes for each fit. The exponents are essentially independent of $p$ as expected, and the slight drifting of the exponent estimated from $\Gamma$ at small $p$ is mainly due to the finite-size effect. The finite-size effect appears to become stronger as $p$ is lowered, presumably because the fluctuation of the graph becomes larger as $p$ is decreased.
%number of neighbours with respect to the mean $\sqrt{pL}/pL=1/\sqrt{pL}$ increases significantly as $p$ vanishes.

The space-filling property and the universal scaling functions are also relevant for the diluted KAS model in the mean-field regime, some typical results are shown in Fig.~\ref{SDRGKAS}. Here, the results are extended to $\sigma=0.2$. 
It might be tempting to view that the space-filling interface of the diluted SK model is merely a consequence of the random graph, however, this perspective is not fully appropriate because the space-filling property is not limited to $\sigma=0$, in line with the full KAS model \cite{Wang:Fractal}. The space-filling property is therefore a result of the sufficiently long-range nature of the interactions, or an effective high dimension of the model. However, stronger finite-size effect may arise, e.g., the finite-size correction appears to be noticeably stronger for $\sigma=0.2$ than that of $\sigma=0$ at $p=0.01$. This makes sense as the former model has stronger bond strength variations than the latter model. Nevertheless, the space-filling feature remains robust. We conclude from our SDRG results that the diluted KAS model has space-filling interfaces in the mean-field regime at least for $0 \leq \sigma \leq 0.2$.

\begin{figure*}
\subfigure[]{\includegraphics[width=0.46\textwidth]{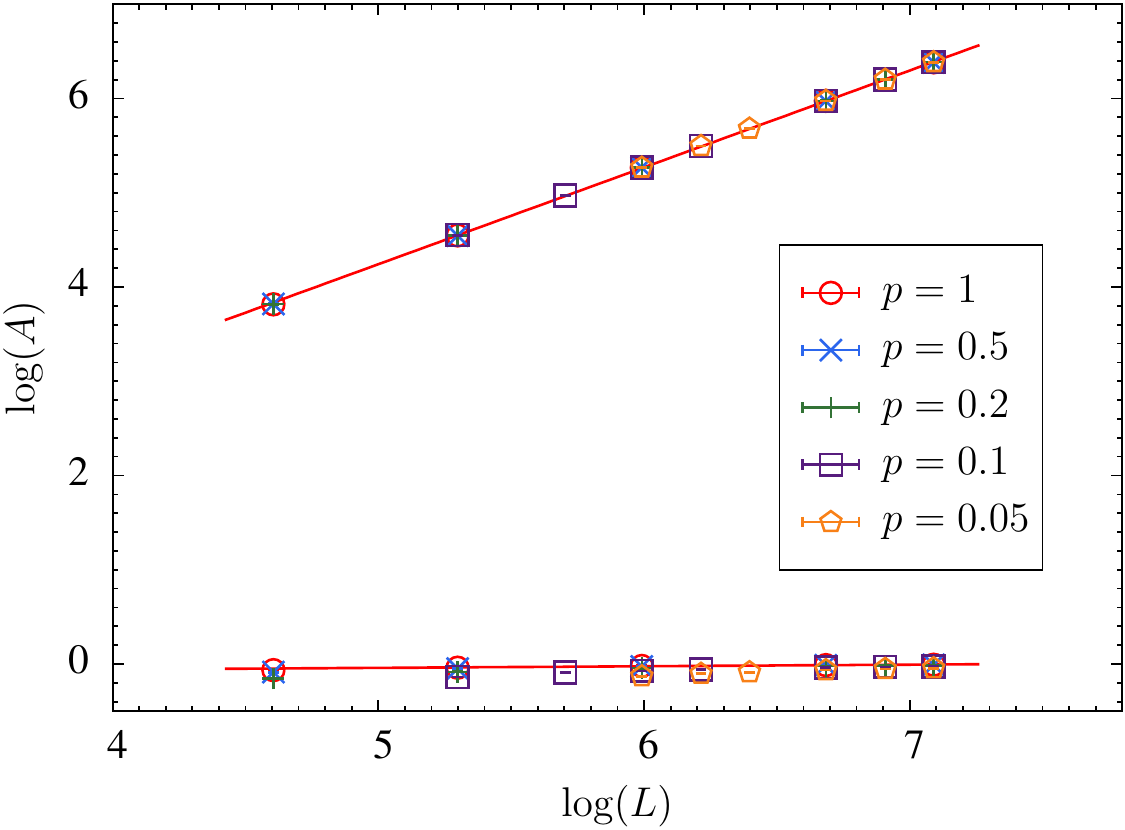} \label{PAMC1}}
\subfigure[]{\includegraphics[width=0.48\textwidth]{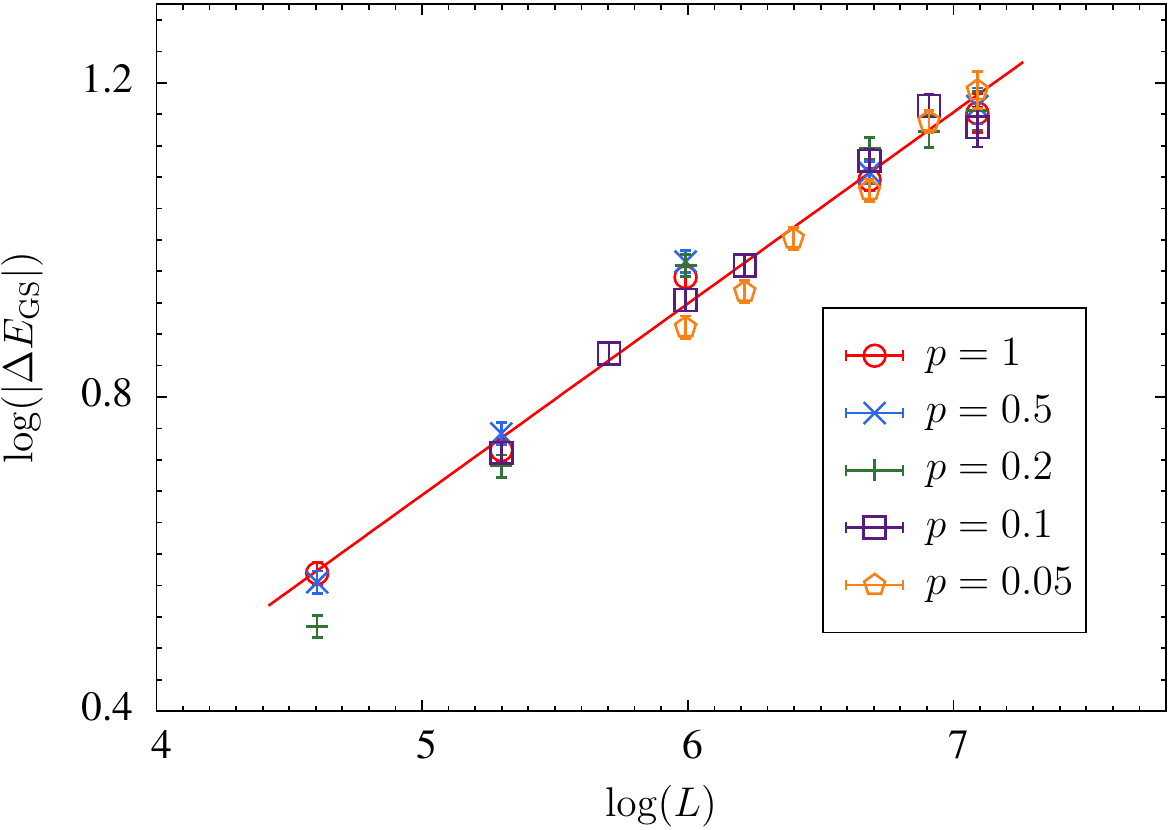} \label{PAMC2}}
\subfigure[]{\includegraphics[width=0.48\textwidth]{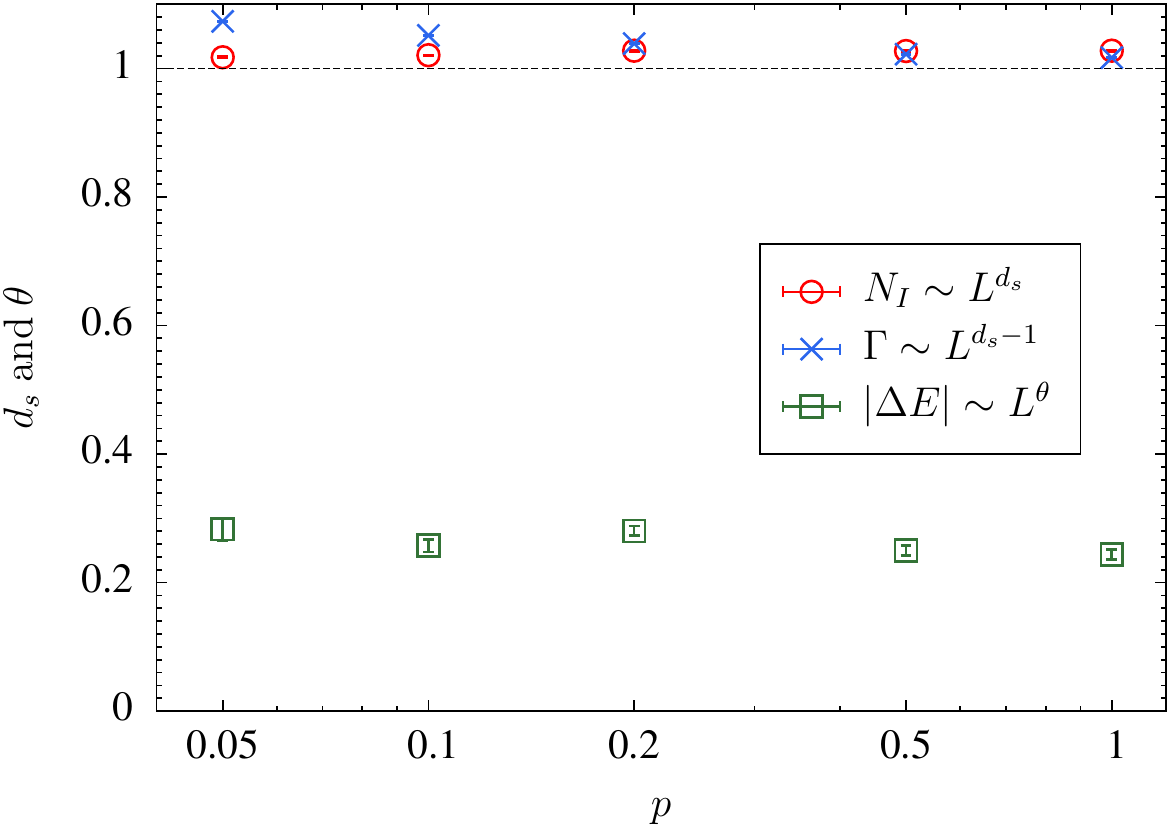} \label{ds2}}
\subfigure[]{\includegraphics[width=0.5\textwidth]{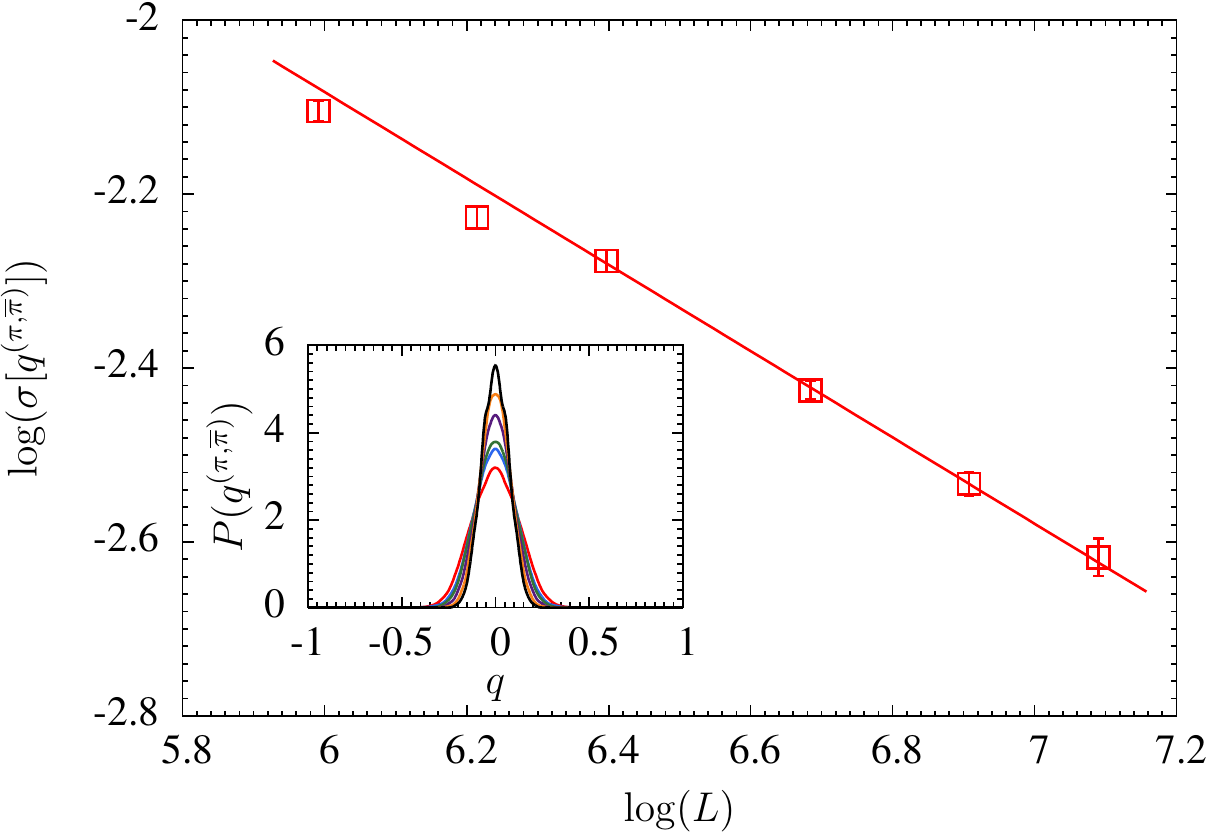} \label{Pqsigma}}
\caption{
The first panel (a) is similar to Fig.~\ref{SDRGSK} but for the PAMC data, and the second panel (b) shows the scaling of the domain-wall energy of Eq.~\eqref{scalingtheta}. The model has strong finite-size effect, but universality appears to exist in $p$. The solid red lines are fits of the $p=1$ data as a guide using all sizes and other fits are omitted here for clarity. The Monte Carlo estimates of $d_s$ and $\theta$ using all sizes are depicted in the third panel (c). The finite-size effect for $d_s$ is considerably larger, but the behaviour is quite similar to that of the SDRG, suggesting that $d_s$ is genuinely a constant. This is further confirmed in the off-diagonal overlap distribution in the final panel (d). The scaling of Eq.~\eqref{sigma} is illustrated at $T=0$ in the main panel, and the fit here uses the four largest sizes. The inset panel depicts that the distribution gets progressively sharper as the system size is increased at $T=0.2$. The panel (c) also suggests that the stiffness exponent $\theta$ is approximately independent of $p$. See the text for more details.
}
\label{PAMC}
\end{figure*}

\begin{table}[t]
\caption{
Simulation parameters of PAMC for the diluted SK model, i.e., at $\sigma=0$. Here, $p$ is bond occupation fraction, $L$ is the system size, $R$ is the population size, $N_T$ is the number of temperatures in the annealing schedule, $N_S$ is the number of Monte Carlo sweeps per replica per temperature, and $M$ is the number of samples. We simulate each sample with both the PBC and the APBC, and the lowest temperature is $T=0.1$.
\label{para2}
}
\begin{tabular*}{\columnwidth}{@{\extracolsep{\fill}} l c c c c r}
\hline
\hline
$p$ &$L$ &$R$ &$N_T$ &$N_S$ &$M$ \\
\hline
$\{1, 0.5, 0.2\}$ &$100$ &$2\times10^4$ &$101$ &$10$ &$3000$ \\
$\{1, 0.5, 0.2,0.1\}$ &$200$ &$5\times10^4$ &$101$ &$10$ &$3000$ \\
$\{1, 0.5, 0.2,0.1,0.05\}$ &$400$ &$1\times10^5$ &$101$ &$20$ &$3000$ \\
$\{1, 0.5, 0.2,0.1,0.05\}$ &$800$ &$2\times10^5$ &$151$ &$20$ &$3000$ \\
$1$ &$1200$ &$2\times10^5$ &$251$ &$20$ &$1034$ \\
$0.5$ &$1200$ &$2\times10^5$ &$251$ &$20$ &$1195$ \\
$\{0.2,0.1,0.05\}$ &$1200$ &$2\times10^5$ &$251$ &$20$ &$1000$ \\
$0.1$ &$300$ &$5\times10^4$ &$101$ &$10$ &$3000$ \\
$\{0.1,0.05\}$ &$500$ &$1\times10^5$ &$101$ &$20$ &$3000$ \\
$0.05$ &$600$ &$2\times10^5$ &$151$ &$20$ &$3000$ \\
$0.2$ &$1000$ &$2\times10^5$ &$201$ &$20$ &$1444$ \\
$\{0.1,0.05\}$ &$1000$ &$2\times10^5$ &$201$ &$20$ &$3000$ \\
\hline
\hline
\end{tabular*}
\end{table}

Because the SDRG method is not exact, we analyze our Monte Carlo data to further validate our SDRG results. This is important for gaining confidence in the SDRG results, despite that the Monte Carlo estimates themselves are not particularly accurate due to the limited sizes.
The corresponding scalings of $N_I$ and $\Gamma$ are shown in Fig.~\ref{PAMC1} and the estimates of $d_s$ using all sizes are depicted in Fig.~\ref{ds2}. The $N_I$ estimates are approximately in the range from $1.0176$ to $1.0278$. The $\Gamma$ estimates are again somewhat worse ranging from $1.0170$ to $1.0734$. In both cases,  using the four largest sizes slightly improve the results. But the improvement is rather marginal due to the small sizes available. Nevertheless, the results are important suggesting that the interfaces are space-filling and the validity of the SDRG results as the data have a qualitatively similar behaviour. The PAMC parameters are summarized in Table~\ref{para2}.

Monte Carlo simulations also provide a compelling evidence of the space-filling nature of the interfaces by studying the off-diagonal overlap distribution. To this end, we introduce the off-diagonal spin overlap as:
\begin{eqnarray}
q^{(\pi, \overline{\pi})} = \frac{1}{L} \sum_i S_i^{(\pi)} S_i^{(\overline{\pi})}.
\end{eqnarray}
This is just the regular spin overlap, except that here two microstates of the PBC and APBC are paired.
Note that this works at both $T=0$ and finite temperatures. In our work, we have collected this overlap distribution at two low temperatures $T=0.2$ and $T=0.1$, both are deep in the spin-glass phase. One important consequence of the space-filling interface is that the off-diagonal overlap distribution should become a delta function $P(q)=\delta(q)$ in the thermodynamic limit. This is a result of the central limit theorem, which can be appreciated from the $O(1)$ islands. Therefore, it is expected that: 
\begin{align}
    \sigma([q^{(\pi, \overline{\pi})}]) &\sim \frac{1}{\sqrt{L}}.
    \label{sigma}
\end{align}

Our data support this sharpening of the off-diagonal overlap distribution, as illustrated in Fig.~\ref{Pqsigma}. By pairing ground states, we find the slope is $-0.462(15)$ for $p=0.05$. The deviation is likely due to the smallest sizes, and by dropping the two smallest sizes, we estimate a slope of $-0.496(29)$, in good agreement with Eq.~\eqref{sigma}. The distribution function is not so smooth, as each sample only contributes one value. Therefore, we illustrate here instead the distribution function at a finite but very low temperature $T=0.2$ in the inset panel. In fact, the temperature here is so low, there is essentially no difference between the $T=0.2$ and $T=0.1$ distributions, except that the latter has stronger fluctuations, the $T=0$ has even stronger fluctuations. It can be seen that the peaks are clearly getting progressively sharper as $L$ is increased, and the distribution of larger sizes starts to take a Gaussian shape. The results are largely similar for $T=0.2$, $T=0.1$, and $T=0$, and also for other $p$ values, and therefore they are omitted here for clarity. By contrast, if $d_s<1$, the winding domain wall will not be system size, and the variance will level off to a finite constant value \cite{Wang:KAS,Wang:DC}.

In summary, our direct SDRG estimates and Monte Carlo estimates of $d_s$ and also the sharpening of the off-diagonal spin overlap distribution strongly suggest that the interfaces of the diluted SK model are space-filling independent of $p$. In addition, the interfaces are likely also space-filling for the diluted KAS model in the mean-field regime independent of both $p$ and $\sigma$ at least for $\sigma \leq 0.2$.

\subsection{Stiffness exponent $\theta$ and $\omega$}

The scaling of the ground state domain-wall energy is shown in Fig.~\ref{PAMC2}. The finite-size effect is quite strong, but all data appear to collapse together, suggesting the existence of a universality in $p$.
The $\theta$ estimates using all sizes are also approximately a constant, ranging from $0.2437$ to $0.2832$ as illustrated in Fig.~\ref{ds2}. Assuming that $\theta$ is indeed independent of $p$, we estimate that $\theta=0.263(5)$ by combining these estimates. If we again using only the four largest sizes, the data are statistically well compatible with a constant but naturally with larger errorbars. It is theoretically expected that $\theta$ is independent of $\sigma$ for the full KAS model in the mean-field regime, the averaged estimate here is close to the result $\theta=0.261(7)$ found at $\sigma=0.1$ with similar ranges of system size \cite{Wang:KAS}. 

\begin{table*}[t]
\caption{
Fits of Eq.~\eqref{ps} using our data, and the comparison with the $\pm J$ results of \cite{Boettcher:SKD}. Our $e_\infty$ fits are fully compatible with the universal value $e_\infty^\star=-0.7631667265(6)$. Therefore, we also estimate $\omega^\star$ and $A^\star$ by fixing $e_\infty=e_\infty^\star$ to improve the accuracy. The results herein suggest that $e_\infty$ is fully universal, the exponent $\omega$ depends on $p$ but not on the disorder, while the numerical factor $A$ depends on both.
\label{para3}
}
\begin{tabular*}{\textwidth}{@{\extracolsep{\fill}} l c c c c c r}
\hline
\hline
$p$ &$1$ &$0.5$ &$0.2$ &$0.1$ &$0.05$ &Disorder \\
\hline
$e_\infty$ &$-0.76323(5)$ &$-0.762(1)$ &$-0.762(1)$ &$-0.762(1)$ &$-0.761(1)$ &$\pm J$ \\
$e_\infty$ &$-0.7639(6)$ &$-0.7637(5)$ &$-0.7628(2)$ &$-0.7629(5)$ &$-0.7625(9)$ &Gaussian \\
$\omega$ &$0.666(3)$ &$0.69(1)$ &$0.73(1)$ &$0.79(1)$ &$0.86(1)$ &$\pm J$\\
$\omega$ &$0.63(3)$ &$0.66(2)$ &$0.76(1)$ &$0.80(2)$ &$0.86(4)$ &Gaussian \\
$\omega^\star$ &$0.666(4)$ &$0.690(4)$ &$0.740(4)$ &$0.784(5)$ &$0.835(7)$ &Gaussian \\
$A$ &$0.71(1)$ &$0.80(4)$ &$1.04(7)$ &$1.7(1)$ &$3.3(5)$ &$\pm J$ \\
$A$ &$0.58(6)$ &$0.75(8)$ &$1.53(7)$ &$2.6(3)$ &$5.6(12)$ &Gaussian \\
$A^\star$ &$0.67(2)$ &$0.84(2)$ &$1.42(3)$ &$2.42(7)$ &$4.9(2)$ &Gaussian \\
\hline
\hline
\end{tabular*}
\end{table*}

The fluctuation in $\theta$ is much smaller than the change in $\omega$ as discussed below, and there is no monotonic trend. Our result therefore suggests that there is not necessarily a strong correlation between $\theta$ and $\omega$, and the formula $\omega=1-\theta$ is likely not correct. This avoids the disturbing consequence $\theta<0$ even if $\omega$ diverges in the $p\rightarrow0$ limit.
Considering that the KAS model and the SK model have strong finite-size effect for the exponent $\theta$ \cite{Wang:KAS}, we conclude that $\theta$ is likely independent of $p$, or alternatively it depends very weakly on $p$ and is approximately independent of $p$.

%It should be a good idea to have better motivations as why we want to know how they vary with $p$.

%In addition, the exponent $\omega$ varies with $p$ in a significant manner, but both the exponents $\theta$ and $d_s$ do not. What this means? There is not necessarily any relation between these exponents.

%We mention in passing that the Gaussian disorder is typically harder to simulate than the $\pm J$ disorder.

Finally, we study the energy scaling. For the full SK model, the average energy per spin is remarkably universal independent of the specific disorder, and the energy is known to great accuracy $e_\infty^\star=-0.7631667265(6)$ \cite{oppermann:07}. To check whether it is also universal in $p$, we do the power series fit of Eq.~\eqref{ps} and the results are summarized in Table~\ref{para3}. Here, we averaged over the PBC and APBC data for each sample for the fit to suppress the statistical error. The result suggests that the fitted $e_\infty$ agrees well with the above ``exact'' result. There is no evidence of a systematic drifting with decreasing $p$. This strongly suggests that the average ground state energy per spin is universal independent of $p$. This is meaningful as the number of bonds diverges and the bond strength is properly normalized, therefore, it is reasonable to expect $e_\infty$ to converge to that of the SK model. The increasing of the fitted $e_\infty$ at very small values of $p$ found in \cite{Boettcher:SKD} is very likely due to that the data therein are affected by including the very small sizes which generate disconnected samples. Here, our minimum size grows with decreasing $p$, such that there are essentially no disconnected samples.

Establishing the universal $e_\infty$ allows us to do a linear fit of $\log(e_L-e_\infty^\star)$ and $\log(L)$ to estimate $\omega$ and $A$ with better accuracy. These estimates are also listed in Table~\ref{para3} as $\omega^\star$ and $A^\star$. The two fits are well compatible, and the linear fit yields much more accurate results. Our data are also compatible with the results of the $\pm J$ model, except for $A$, showing that $e_\infty$ is universal in both $p$ and the disorder, $\omega$ is universal in the disorder but not in $p$, and the prefactor $A$ has no universality.

We comment that $\omega$ does not change very dramatically down to $p=0.05$, it is interesting whether the rapid divergence at yet smaller $p$ values \cite{Boettcher:SKD} is a genuine divergence or finite-size effect. The sizes in the pioneering work \cite{Boettcher:SKD} are quite small for the small $p$ values, the samples are likely quite disconnected without suitably tuning the minimum $L$ as $p$ is lowered. Future work should double check this at yet smaller $p$ values.
%However, this can be difficult to check as a smaller $p$ requires yet larger sizes.

\section{Conclusions and Future Challenges}
\label{conclusion}
In this work, we studied the interface properties of the diluted SK model using SDRG and PAMC simulations. We find that the interfaces are space-filling with $d_s=1$ and the stiffness exponent $\theta$ is also approximately a constant independent of $p$. These are also likely true for the diluted KAS model upto at least $\sigma=0.2$. The energy finite-size scaling is also studied and compared with that of the $\pm J$ disorder, finding that the thermodynamic energy is universal in both $p$ and the disorder, the exponent $\omega$ is universal in the disorder but not in $p$, while the prefactor has no universality.

Assuming the diluted SK model and the full SK model have the same properties of interest, this work provides an opportunity to accelerate simulating mean-field spin glasses. This is because it can be much faster to update a sample if the connectivity is considerably lowered, e.g., $p=0.1$. However, one also needs to consider that the minimum size grows and the fluctuation can also grow at small $p$, therefore, a balance should be kept between the two effects.

Our work can be extended in a number of directions. First, it should be interesting to study these exponents more systematically in the full range of $\sigma$ including the droplet regime \cite{Wang:Fractal}. As mentioned earlier, a large-scale simulation at yet smaller $p$ values to check the divergence of $\omega$ is also interesting. Another related research direction in our setup is the study of the SK model or the diluted SK model in thermal boundary conditions and examine the quantity of the sample stiffness \cite{Wang:TBC}. This is of paramount importance for determining whether the three-dimensional Edwards-Anderson model has a single pair or many pairs of pure states. Research efforts along these research directions are currently in progress, and will be reported in future publications.

\begin{acknowledgments}
We gratefully acknowledge supports from the National Science Foundation of China under Grant No. 12004268, the Fundamental Research Funds for the Central Universities, China, and the Science Speciality Program of Sichuan University under Grant No. 2020SCUNL210. We thank the Emei cluster at Sichuan university for providing HPC resources.
\end{acknowledgments}

\bibliography{Refs}

\end{document}